\documentstyle[12pt,epsf,world_sci]{article}
\begin{document}
\newcommand{\abseta}{\mid \eta \mid \leq}
\newcommand{\ptran}{{\rm P}_{\scriptscriptstyle\rm T}}
\begin{flushright}
\vspace*{-0.6in}
%                                                CDF/PUB/BOTTOM/PUBLIC/2739 \\
                                                 FERMILAB-CONF-94/328-E    \\
                                                       Submitted to DPF '94 \\
\end{flushright}

\title{MEASUREMENT OF CORRELATED $b$ QUARK CROSS SECTIONS AT CDF}
\author{DAN AMIDEI, PAUL DERWENT, DAVID GERDES, and TAE SONG\thanks{Presented
        by David Gerdes, representing the CDF Collaboration.}\\
        {\em Randall Laboratory of Physics, University of Michigan, \\
             500 E. University Avenue, Ann Arbor, MI 48109} \\ }
\maketitle
\setlength{\baselineskip}{2.6ex}

\begin{center}
\parbox{13.0cm}
{\begin{center} ABSTRACT \end{center}
{\small \hspace*{0.3cm} Using data collected during the
1992-93 collider run at Fermilab,  CDF has
made measurements of correlated $b$ quark cross section where one $b$ is
detected via a muon from semileptonic decay and the second $b$ is detected
with secondary vertex techniques. We report on measurements of the cross
section as a function of the momentum of the second $b$ and as a function of
the azimuthal separation of the two $b$ quarks, for transverse momentum of the
initial $b$ quark greater than 15 GeV. Results are compared to QCD
predictions.}}
\end{center}

\par Studies of $b$ production in $p\bar{p}$ collisions provide
quantitative tests of perturbative QCD.  Measurements of the cross section for
$p\bar{p}\rightarrow bX$ have been made at CDF~\cite{latest_CDF} and
UA1~\cite{old_UA1}. In this analysis we extend the range of comparisons
between theory and experiment by performing a measurement of $b$-$\bar{b}$
correlations in the process $p\bar{p}\rightarrow b\bar{b}X$.
We identify the first $b$ via its semileptonic decay to a muon,
and the other $b$ (referred to for simplicity as the $\bar{b}$, though we do
not
perform explicit flavor identification for either $b$)
by using precision track reconstruction to measure the displaced tracks
from $\bar{b}$ decay. Identification of the $b$ and $\bar{b}$ permits
a measurement of the cross section as a function of the transverse
momentum of the $\bar{b}$, $\frac{d\sigma_{\bar{b}}}{dE_T}$,
and as a function of the azimuthal
separation of the two $b$ quarks, $\frac{d\sigma_{\bar{b}}}{d\delta\phi}$,
for transverse momentum of the initial $b$ quark greater than 15 GeV.
The data used here were collected by the Collider Detector at Fermilab (CDF)
during the 1992-93 Tevatron collider run, and correspond to an integrated
luminosity of 15.1$\pm$0.5~pb$^{-1}$.

\par The CDF has been described in detail elsewhere~\cite{NIM_book}.
The tracking systems used for this analysis are the silicon vertex
detector\cite{SVX_nim} (SVX), the
central tracking chamber (CTC), and the muon system. The central
muon system consists of two detector elements.  The Central Muon chambers
(CMU) provide muon
identification over 85\% of $\phi$ in the pseudorapidity range
$|\eta|\leq$0.6, where $\eta = - \ln[\tan(\theta/2)]$.  This $\eta$ region is
further instrumented by the Central Muon Upgrade chambers (CMP), located
behind the CMU after an additional $\approx$
3 absorption lengths.  The calorimeter systems used for this analysis are the
central and plug systems, which give 2$\pi$ azimuthal coverage in the range
$|\eta|< $ 1.1 and $1.1 < |\eta| < 2.4$ respectively.

{}From events that pass an inclusive muon trigger, we select good-quality
CMU muons with $P_T > 9$~GeV that have an associated track segment
in the CMP. We further require the muon track to fall within the fiducial
region of the SVX. This sample contains 145,784 events. An independent
analysis has measured the fraction of muons from $b$-decay in this sample
to be 36.0$\pm$2.4$\pm$2.5\%~\cite{Tae}, making this an excellent sample
to look for the presence of additional $\bar{b}$ jets.

The long lifetime of $b$ quarks causes the tracks from $b$-decay
to be displaced relative to the primary $p\bar{p}$ interaction point.
The high-precision track reconstruction made possible by the SVX allows
us to identify these tracks with good efficiency. We use the ``jet-probability"
algorithm~\cite{jetprb}, which compares the impact parameters of the
tracks in a jet to the measured resolution of the SVX
and determines an overall probability that the jet is primary. This probability
is flat between 0 and 1 for jets from zero-lifetime particles, and has
a peak at low values for jets from $b$ and $c$ decays.

To identify the $\bar{b}$ jets in the inclusive muon sample, we require
that the event contain at least one jet with
$E_T>$ 10 GeV, $|\eta|$ $<$ 1.5, and
at least 2 good tracks. The jet is required to be separated from the muon in
$\eta-\phi$ space by $\Delta$R $\geq$ 1.0, so that
the tracks clustered around the jet axis are separated from the
$\mu$ direction. There are 17810 events in this sample.
We then fit the jet probability distribution of these
jets to a sum of Monte Carlo templates for $b$, $c$, and primary jets,
thereby obtaining the fraction of $\bar{b}$ jets in the sample.

The Monte Carlo samples for $b$ and $c$ jets are produced using
the ISAJET event generator\cite{isajet} and a full detector simulation.
The CLEO Monte Carlo program~\cite{qq} is used to model
the decay of $B$ mesons, using an average $b$ lifetime of
$c\tau$ = 420 $\mu$m. The input jet probability templates for
$b$, $c$, and zero-lifetime jets are shown in Figure~1a. The fit is actually
done using the variable $\log_{10}$(jet probability), which magnifies the
interesting region of low jet probability.
Tests of the fitter in Monte Carlo samples with known
admixtures of $b$, $c$, and primary jets show that the fitter
returns the correct number of jets of each type to within the uncertainties.
The data are shown
in Figure~1b together with the results of the fit. The fit agrees very well
with
the data over ten orders of magnitude in jet probability,
and predicts 2620 $\pm$ 97 $\bar{b}$ jets, 2085
$\pm$ 180 $c$ jets, and 13103 $\pm$ 161 primary jets for a total of 17808.
\begin{figure} % Note the following piece of Tex artistry
\begin{minipage}[t]{3in}
   \epsfxsize=3in
   \epsfysize=3in
   \epsfbox[0 162 500 657]{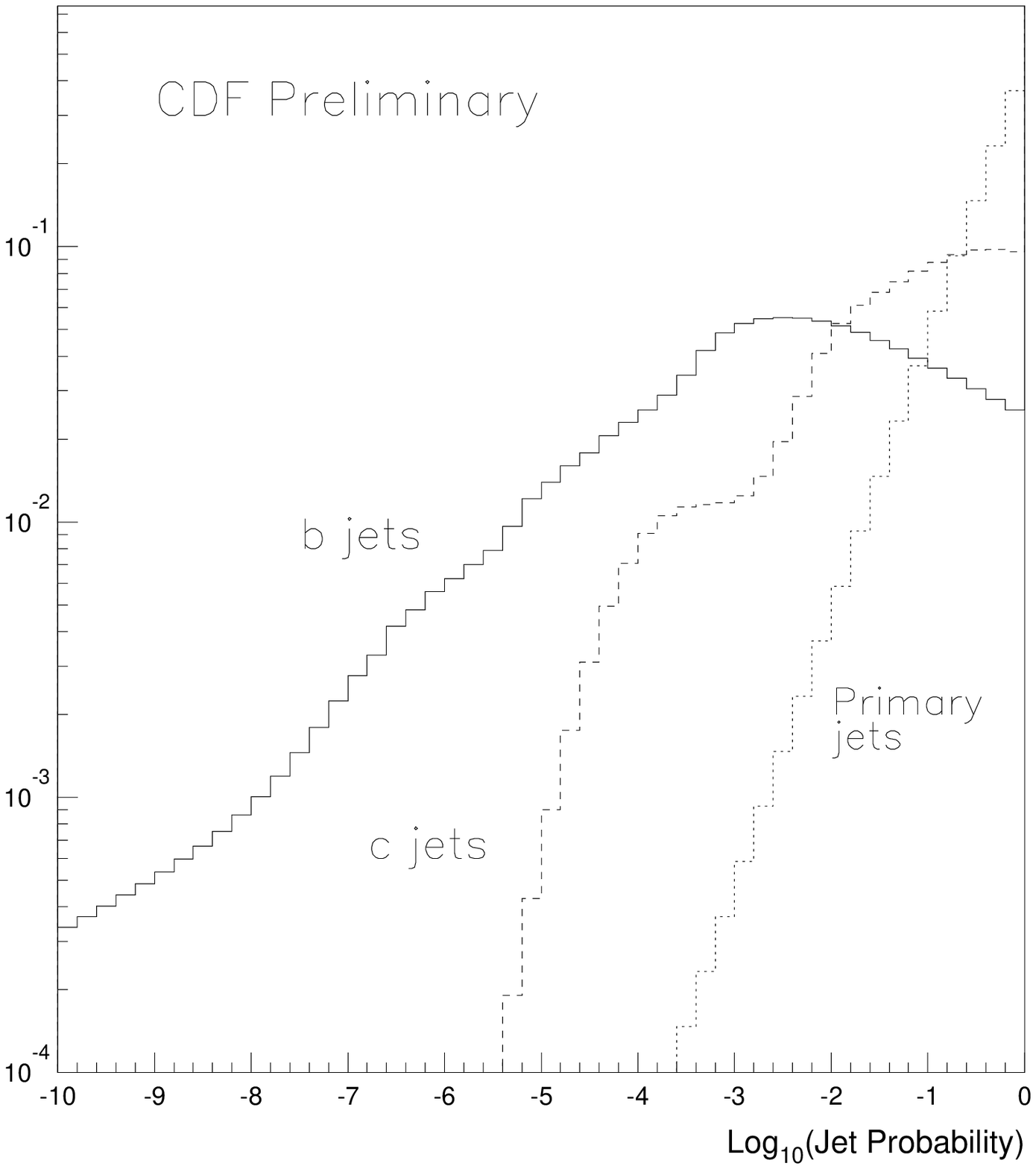}
\end{minipage} \ \
\begin{minipage}[t]{3in}
   \epsfxsize=3in
   \epsfysize=3in
   \epsfbox[0 162 500 657]{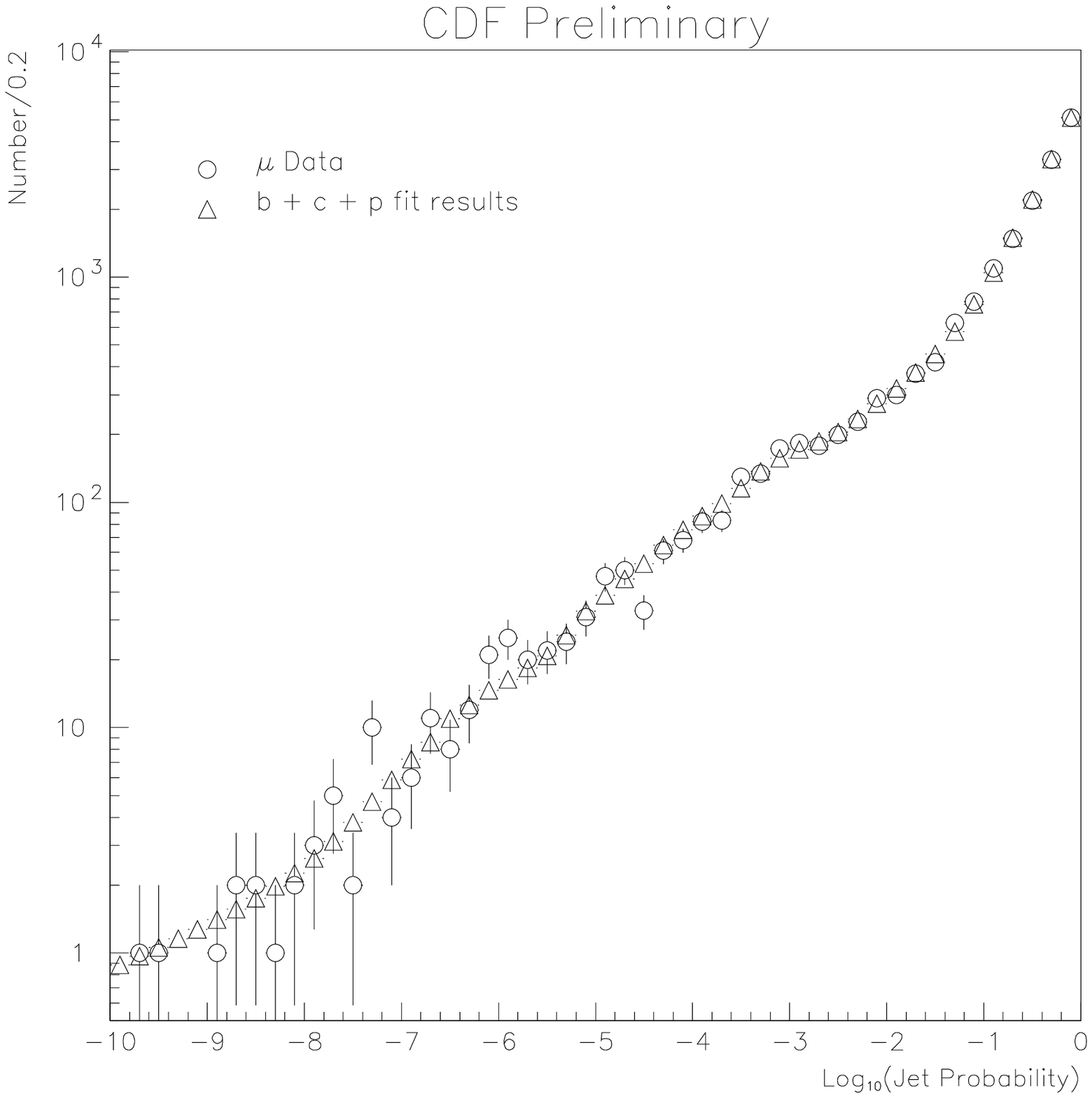}
\end{minipage}
\caption{(a) Jet probability distributions for $b$, $c$, and primary
jets. (b) Observed jet probability distribution for jets in the inclusive
muon sample, together with the results of the fit.}
\end{figure}

To convert the results of the above fit into a measurement of the
cross section, we calculate the acceptance for both $b\rightarrow\mu$
and $\bar{b}$ jets.
The $\mu$ acceptance and efficiency has three parts: (1) the
fiducial acceptance for muons coming from $b$'s with $|\eta|<1$, (2) the
fraction of $b$'s, $P_T^b > P_T^{min}$, that decay to muons with $P_T>$ 9 GeV,
and (3) the trigger and identification efficiencies for 9 GeV muons. The first
two factors are studied using the Monte Carlo sample described above.
The $P_T^{min}$ value, chosen
such that 90\% of muons with $P_T >$ 9 GeV come from $b$ quarks with
$P_T > P_T^{min}$, is 15~GeV.
The trigger and identification efficiencies for muons
are determined from J/$\psi$ and Z$^{\circ}$ samples.  The overall
acceptance for $b\rightarrow\mu (P_T^{\mu}>9 {\rm GeV}, P_T^b > 15~{\rm GeV},
|y^b|<1)$, including the semileptonic branching ratio\cite{CLEO_br}, is
$0.239^{+0.030}_{-0.018}$\%.

\par The $\bar{b}$ jet acceptance represents the fraction of $\bar{b}$ quarks
that produce jets with
$E_T >$ 10 GeV, $|\eta| <$ 1.5 and at least 2 good tracks inside a cone of 0.4
around the jet axis, in events where there is also a $b$ quark
which decays to a $\mu$
with $P_T >$ 9 GeV within the CMU-CMP acceptance. The $\bar{b}$ jet
acceptance is calculated separately as a function of the jet $E_T$ and
azimuthal opening angle between the two quarks, using the Monte Carlo
sample described above.
The average acceptance for the $\bar{b}$ is $\approx$~40\%, and
ranges from 32.9 $\pm$ 1.9\% (statistical error only) for 10 $< E_T <$ 15
GeV to 49.8 $\pm$ 7.3\% for 40 $< E_T<$ 50 GeV.  For $\delta\phi <
\frac{\pi}{8}$ radians, the acceptance is 7.3 $\pm$ 2.2\%, while for
$\frac{7\pi}{8} < \delta\phi < \pi$, the acceptance is 51.4 $\pm$ 0.8\%.

\par  We have compared the values for the $\bar{b}$ jet acceptance from
ISAJET samples to the acceptance from HERWIG samples.  The acceptance agrees
within the statistical error in the samples as a function of $E_T$, differing
at the 5\% level.  We take this as an additional systematic uncertainty on
the acceptance.  In combination with a 10\% uncertainty due to the vertex
distribution for events in the SVX fiducial volume, we have a common  11.2\%
systematic uncertainty in all the jet acceptance numbers.

We use the jet probability fit to determine the number of $\bar{b}$
jets as a function of (1) the azimuthal
separation $\delta\phi$ between the jet and the muon,
and (2) the $E_T$ of the jet, and convert
these numbers into the differential cross section using the aceptances
calculated above.
Figure~2a shows the measured distribution of
$d\sigma_{\bar{b}}/d(\delta\phi)$, together with
a prediction from the Mangano-Nason-Ridolfi (MNR) calculation~\cite{MNR}.
There is a large change in the
acceptance for $\delta\phi < \frac{3\pi}{8}$ due to the
$\Delta$R separation requirement on the $\mu$--jet system.
The shapes of the
theoretical prediction and the experimental data agree well, especially for
$\delta\phi > \frac{\pi}{2}$, but the overall normalization of the
data is about a factor of 1.3 higher than predicted.

We also divide the jets into six $E_T$ bins between 10 and 50 GeV,
and fit the jet probability distribution for each bin to determine
$d\sigma_{\bar{b}}/dE_T$.
This cross section is shown in Figure~2b.
Work to compare this measurement to the MNR prediction, which requires
a bin-by-bin understanding of the detector response, is in progess.
\begin{figure}
\begin{center}
\epsfysize=2.6in
\leavevmode\epsfbox[45 414 500 657]{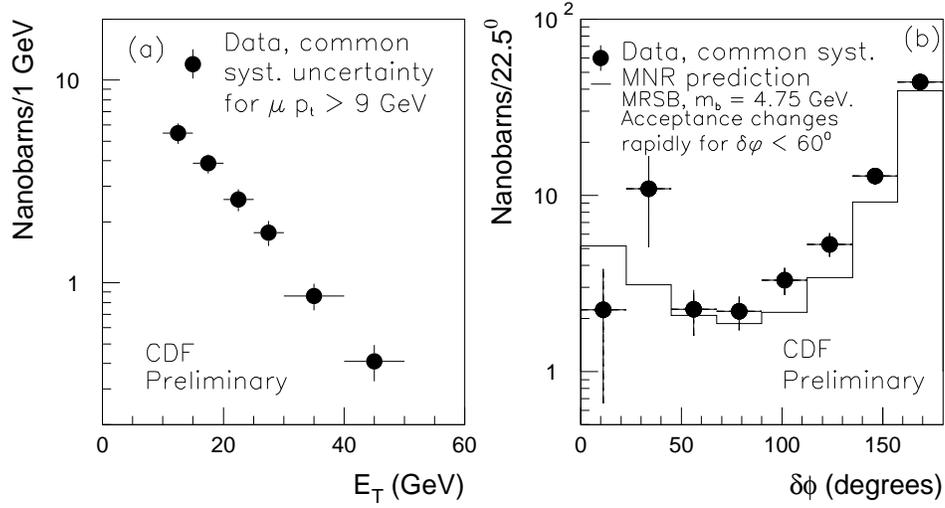}
\caption{Measured differential cross section as a function
of (a) the $E_T$ of the
$\bar{b}$ jet, and (b) the azimuthal angle between the jet
and the muon, for jets with $E_T>10$~GeV and $|\eta|<1.5$.
There is a common systematic uncertainty of
$^{+18.3}_{-15.4}$\% not shown in the experimental points. Also
shown in (b) is the prediction from the MNR calculation.}
\end{center}
\end{figure}

\par We thank the Fermilab staff and the technical staffs of the
participating institutions for their vital contributions.  This work was
supported by the U.S. Department of Energy and National Science Foundation,
the Italian Istituto Nazionale di Fisica Nucleare, the Ministry of Education,
Science and Culture of Japan, the Natural Sciences and Engineering Research
Council of Canada, the National Science Council of the Republic of China,
the A. P. Sloan Foundation, and the Alexander von Humboldt-Stiftung.

\end{document}